\documentclass[aps,twocolumn,showpacs,floats,superscriptaddress]{revtex4}
\usepackage{graphicx}
\usepackage{amssymb,amsmath}
\usepackage{color}
\begin{document}

\author{L. Pollet}
\affiliation{Department of Physics, University of Massachusetts,
Amherst, MA 01003, USA}
\author{N.V. Prokof'ev}
\affiliation{Department of Physics, University of Massachusetts,
Amherst, MA 01003, USA}
\affiliation{Russian Research Center ``Kurchatov Institute'',
123182 Moscow, Russia}

\author{B.V. Svistunov}
\affiliation{Department of Physics, University of Massachusetts,
Amherst, MA 01003, USA}
\affiliation{Russian Research Center ``Kurchatov Institute'',
123182 Moscow, Russia}

\author{M. Troyer}
\affiliation{Theoretische Physik, ETH Zurich, 8093 Zurich, Switzerland}

\title{Absence of a Direct Superfluid to Mott Insulator Transition in Disordered Bose Systems}

\date{\today}
\begin{abstract}
We prove the absence of a direct quantum phase transition
between a superfluid and a Mott insulator in a bosonic system with generic, bounded disorder.
We also prove compressibility of the system on the superfluid--insulator critical line
and in its neighborhood. 
These conclusions follow from a general  {\it theorem of inclusions} which states that for any 
transition in a disordered system one can always find rare regions of the competing phase 
on either side of the transition line.
Quantum Monte Carlo simulations for the disordered Bose-Hubbard model 
show  an even stronger result, important  for the nature of the Mott insulator to Bose glass phase transition:
The critical disorder bound, $\Delta_c$, corresponding to the onset of
disorder-induced superfluidity, satisfies the relation $\Delta_c > E_{\rm g/2}$,
with $E_{\rm g/2}$ the half-width of the Mott gap in the pure system. 
\end{abstract}



\maketitle


The interplay between disorder and interactions is one of the most exciting
and longstanding problems in condensed matter physics, full of controversial theoretical
and experimental results. Bosonic systems are especially difficult to handle theoretically
because the limit of vanishing interactions is pathological: in the ground state all particles
occupy the same lowest-energy localized state. Hence, interactions and disorder are equally important
at the quantum phase transition between superfluid (SF) and insulating (I)
ground states. Relevant examples from Nature include ${}^4$He in porous media and aerogels~\cite{Reppy}, ${}^4$He on substrates~\cite{Reppy,Chan}, thin superconducting films~\cite{Goldman}, and Josephson junction arrays~\cite{Vanderzant}.

The physics is further complicated when bosons are subject to a periodic external potential
and interactions are strong enough to drive a commensurate system (with particle number an integer multiple of the number of lattice sites) to the insulating Mott phase (MI) even without disorder. A fundamental problem here is  the role of an arbitrarily weak disorder in the vicinity of SF-MI quantum critical point of the pure system.

Building on one-dimensional results by Giamarchi and Schulz \cite{GS} Fisher {\it et al.} 
argued the existence of a {\it Bose glass} phase (BG) in the disordered Bose-Hubbard Hamiltonian in any dimension
 \cite{Fisher}: while the commensurate MI is gapped, the insulating BG remains compressible.
Ever since, the Bose-Hubbard model has received a lot of theoretical attention,
although it remained beyond direct experimental reach.
This is changing with the experimental demonstration of the SF-MI transition of ultra-cold atoms in optical lattices~\cite{Greiner} following the theoretical proposal of Ref.~\cite{Jaksch}.
At present, cold-atom systems offer an unprecedented control over system properties and have been explicitly shown to be accurately described by the Bose-Hubbard model. In particular,
there is full agreement between the experimental data and Quantum Monte Carlo simulations,
proving that the optical lattice system can be a quantitatively reliable {\it simulator} \cite{Trotzky}.
Experiments have recently advanced to the stage where disorder can be added and controlled using speckle potentials \cite{Aspect,Inguscio,White}.

Though the pure system is well understood, the properties of the disordered Bose-Hubbard model
remain subject to debate, even at the qualitative level. The relevance arbitrarily weak disorder at the superfluid--insulator quantum phase transition at an integer filling factor is a highly
controversial issue \cite{Fisher,FM96,Scalettar,Krauth,Zhang,Singh,Makivic,Wallin,Pazmandi95,Pai,Svistunov96,Pazmandi98,Kisker,Herbut,Sen,LeeChaKim,PS_04,Wu_Phillips,Bissbort,Weichman1,Weichman2}.
Fisher {\it et al.}~\cite{Fisher} raised the problem and argued that a direct SF-MI transition
was unlikely (see also Refs.~\cite{Weichman1,Weichman2}), though not fundamentally impossible. Curiously, a large number of direct numerical simulations~\cite{Krauth,Makivic,Wallin,Pai,Sen,LeeChaKim} and some approximate approaches \cite{Zhang,Singh,Pazmandi95,Pazmandi98,Wu_Phillips,Bissbort} observed this unlikely scenario!  \\

However,
{\it all} numerical simulations reporting a direct SF-MI transition were ignoring a rigorous theorem
\cite{Fisher,FM96} (to be referred to as {\it Theorem I}) stating that if the bound $\Delta$ on the
disorder strength is larger than the half-width of the energy gap $E_{\rm g/2}$ in the ideal
Mott insulator, then the system is inevitably compressible, {\it i.e.} the transition is to the
BG insulator and not the MI whenever
\begin{equation}
\Delta_c > E_{\rm g/2} .
\label{condition}
\end{equation}

 Intuitively, the condition $\Delta > E_{\rm g/2}(U)$ seems to be necessary for the disorder to at least destroy the Mott gap, putting aside the question of onset of superfluidity which is likely to require even stronger disorder. Therefore, it is reasonable to conjecture that  Eq.~(\ref{condition}) holds for the Bose-Hubbard model (see Refs.~\cite{Weichman1,Weichman2} for more details). This condition was already shown to be necessary in 1D \cite{Svistunov96}, and there is compelling numerical evidence that it equally holds in 2D \cite{PS_04}. Nevertheless, a rigorous proof valid for any dimension has not been found. Moreover, it was recently  claimed \cite{Wu_Phillips} that the relevance of weak disorder depends on dimensionality and that in 3D weak disorder can be irrelevant. 

In this Letter, we introduce and  prove the {\it theorem of inclusions}: In 
the presence of generic bounded disorder  there exist rare, but arbitrarily large,
regions of the competing phase across the transition line. By generic disorder we mean that any  particular realization  
has a non-zero probability density to occur in a finite volume. This theorem immediately implies  the absence of a direct SF-MI quantum phase transition.  

We will also introduce {\it Theorem II} which states a non-zero compressibility  on the
superfluid--insulator critical line and in its neighborhood for models which have disordered on-site
potentials.  {\it Theorem II} is not based on, and thus {\it does not imply} condition (\ref{condition}).
To check whether Eq.~(\ref{condition}) holds in three dimensions, as well as to produce some
quantitative benchmarks for future experiments, we complement the theorem by first-principles simulations of the disordered Bose-Hubbard model by the worm algorithm \cite{worm}. 
For weak disorder, our results are generic through standard long-wavelength universality considerations.

The Hamiltonian of the Bosonic Hubbard model with on-site disorder reads:
\begin{equation}
H = -t \! \sum_{\langle i,j\rangle}\! a_i^{\dagger} a_j + {U\over 2} \sum_i n_i (n_i-1) + \sum_i ( \varepsilon_i -\mu) n_i .
\label{Bose_Hubbard}
\end{equation}
The subscripts label the sites of a (simple cubic) lattice, $a_i^{\dagger}$ is the boson creation operator,
the symbol $\langle i,j\rangle$ denotes nearest-neighboring sites, $n_i=a_i^{\dagger} a_i$ is the density operator. In what follows we use the amplitude of the hopping term $t$ as the unit of energy,
$U$ is the on-site pair interaction, $\mu$ is the chemical potential, and
$\varepsilon_i$ represents the random disorder potential. Normally,
one assumes---without loss of generality---that there
are no correlations between the $\varepsilon_i$'s on different sites,
and that their values are uniformly distributed on the $[-\Delta, \Delta ]$ interval.

The proof of Theorem I is based on the fact that in the infinite system one can always find arbitrary large `Lifshitz' regions where the chemical potential
is nearly homogeneously shifted downwards or upwards by $\Delta$. There is no energy gap for particle transfer between such regions, and they
can be doped with particles/holes \cite{conjecture}.
Note also that Theorem I immediately implies that the MI phase does not exist in a system
with unbounded disorder.

This proof gives a hint of a potential pitfall when one tries to directly interpret numerical or experimental data on finite system sizes.
For $\Delta \to 0$ the distance between regions contributing to a non-zero, but {\it exponentially} small, compressibility is diverging {\it exponentially}. If disorder is not very strong,
a typical finite-size cluster is not supposed to contain even a single rare region,
and the system behavior perfectly mimics a direct SF-MI transition.
At the same time, Theorem I also provides a straightforward way out. One just needs to compare the critical
amplitude of the disorder, $\Delta_c(U)$, to $E_{\rm g/2}(U)$: If condition (\ref{condition})
is satisfied, then the transition is towards the Bose glass rather than the Mott insulator, with the compressibility remaining {\it finite} at the critical point. We now proceed to the proof of the theorem of inclusions.

 {\it Proof of the theorem of inclusions.}  The proof is straightforward if reducing the disorder bound enhances the superfluid phase (cf. e.g. Refs.~\cite{Harris, ImryWortis}). In the opposite case, let us introduce $\vec{\xi}$ denoting all microscopic parameters other than the bound $\Delta$ that fix the shape and correlations of the disorder distribution. In the parameter space  $(U,\Delta,\vec{\xi}\, )$, the critical hyper-surface of the superfluid to insulator quantum phase transition is written as $\Delta = \Delta_c(U,\vec{\xi}\, )$. For simplicity,
we keep $U$ fixed, without loss of generality. On physical grounds, $\Delta_c(\vec{\xi}\, )$ is a continuous
function of its microscopic parameters, and a  generic $\vec{\xi}\,$
is not an extremum of $\Delta_c$. We do not require that $\Delta_c(\vec{\xi}\, )$ is analytic,
even though this is most likely to be the case physically. Generic disorder implies that there exist arbitrarily large regions in which the disorder realization generated by $\vec{\xi}$ can be considered with finite probability density as a typical realization of a different set $\vec{\xi}_{\rm fluct}$. In a sufficiently small neighborhood of the critical surface, we can always find  $\vec{\xi}_{\rm fluct}$ such that all points $\Delta \in [ \Delta_c(\vec{\xi}\, ),\Delta_c(\vec{\xi}_{\rm fluct})]$ lie in a superfluid domain of the phase diagram thanks to $\vec{\xi}$ being non-extremal.  The deviation from the critical surface, $\delta \Delta = \Delta_c(\vec{\xi}_{\rm fluct})-\Delta_c(\vec{\xi}\,) $, can be chosen small enough to guarantee that it can be compensated by changing  $\vec{\xi}$ to $\vec{\xi}_{\rm fluct}$.
We conclude that within a distance $\vert \delta \Delta \vert$ from the critical surface the system develops arbitrarily large superfluid domains, which, by virtue of standard Lifshitz tail arguments 
implies the absence of the gap (as in Theorem I) and thus the absence of a direct SF-MI transition.

{\it Proof of theorem II.} 
Close enough to the critical surface we can choose a small finite value $\delta \Delta < 0 $  since fluctuations of $\vec{\xi}$ can always be used to compensate for a sufficiently small deviation from the superfluid domain. In such a rare superfluid domain, the chemical potential can  be homogeneously shifted by a finite amount $\delta \mu \sim  \delta \Delta $ without violating the global disorder bound. Since one can always find rare superfluid regions large enough such that the finite-size quantization of the energy is smaller than the allowed variation $\delta \mu$, the density of states at the global chemical potential is guaranteed to be finite, {\it i.e.} the system is compressible across the SF-I transition. Note that our proof is not based on a specific form of the Hamiltonian and thus applies to all bosonic systems with the on-site potential disorder featuring SF-I quantum phase transition.

\begin{figure}
\centerline{\includegraphics[angle=-90, width=0.9\columnwidth]{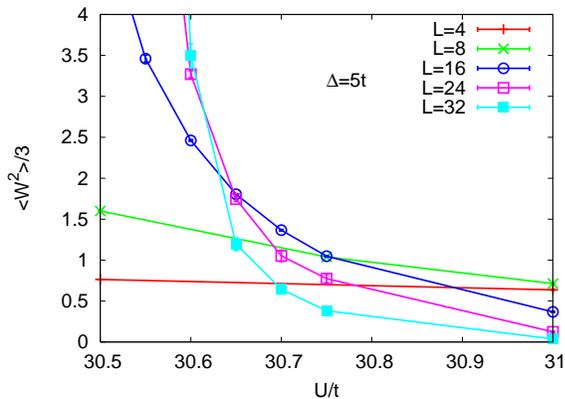}}
\caption{ (Color online). Winding number squared as a function of interaction strength $(U/t)$ for $\Delta = 5t$ using a scaling analysis with the temporal exponent $\tilde{z}=2$, chosen for numerical convenience (see text). The corresponding pairs of the linear system size and inverse temperature are : $L=4$ and $\beta t = 1$, $L=8$ and $\beta t = 4$, $L=16$ and $\beta t = 16$, $L=24$ and $\beta t = 36$, and finally $L=32$ and $\beta t = 64$. The values of the chemical potentials were taken from Ref.~\cite{BCS}. }
\label{fig:sf5}
\end{figure}

\begin{figure}
\centerline{\includegraphics[angle=-90, width=0.9\columnwidth]{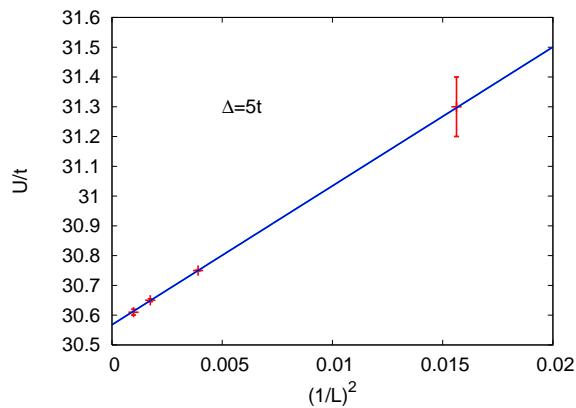}}
\caption{ (Color online). Finite size scaling of critical points $(U/t)_c(L)$ as a function of $1/L^2$ for disorder strength $\Delta = 5t$. The data points shown here are the intersection points in Fig.~\ref{fig:sf5}. In the thermodynamic limit we find $(U/t)_c = 30.57 \pm 0.02$ by extrapolating according to $(1/L)^2$ (see text). }
\label{fig:intersect5}
\end{figure}

\begin{figure}
\centerline{\includegraphics[angle=-90, width=0.9\columnwidth]{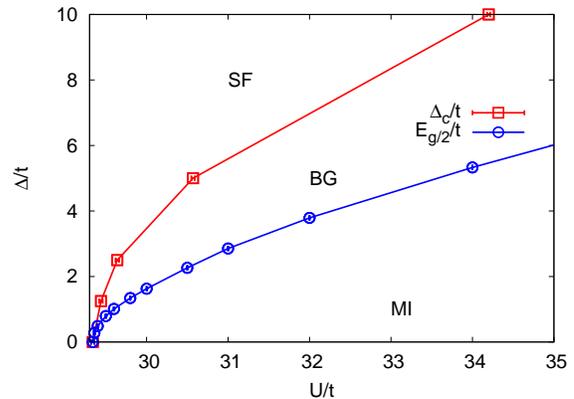}}
\caption{(Color online). Phase diagram of the 3D Bose-Hubbard model with disorder.
The $E_{g/2}(U)$ curve is taken from Ref.~\cite{BCS} and is shown here to demonstrate that
the condition $\Delta_c(U) > E_{g/2}(U)$ is definitely satisfied. It also marks the
BG-MI transition boundary according to the conjecture.
Error bars are shown, but are smaller than point sizes. }
\label{fig:phasediagram}
\end{figure}

{\it Numerical analysis.}
In our numerical study of model (\ref{Bose_Hubbard}) we consider
$\varepsilon_i$ uniformly distributed on the $[-\Delta, \Delta]$ interval, and choose the chemical potential in the middle of the $n=1$ Mott lobe \cite{BCS} to ensure that the system is always in the $\langle n \rangle =1$ state. We employ the worm algorithm, implemented in two independent ways~\cite{worm,worm_lode}. Our finite-size scaling needed for the SF-I transition is based on the square of the winding number of world lines $\langle W^2 \rangle$, related to the superfluid stiffness by $\Lambda_s=\langle W^2 \rangle /TLd$~\cite{Pollock}.
In Fig.~\ref{fig:sf5} we show $\langle W^2 \rangle$ for different values of $U/t$ and system sizes at fixed disorder bound $\Delta = 5t$ where we performed scaling with the temporal exponent $\tilde{z}=2$ which is our choice to reduce the numerical effort. The intersection points of the curves with different system sizes, are shown in Fig.~\ref{fig:intersect5}. Extrapolation to the limit of infinite system size  using a $(1/L)^2$ scaling yields a critical value
$(U/t)_c = 30.57(2)$. Although the exponents for corrections to scaling are not known precisely, the critical values $(U/t)_c$ are determined very accurately; we have repeated the
simulations with $\tilde{z}=1$ (not shown) and found the same value of $(U/t)_c$ within error bars, showing that the choice of $\tilde{z}$ does not matter for the location of the critical point. Indeed, winding numbers are exponentially suppressed in the insulating phase, and thus
any temporal exponent is supposed to yield the same critical point in the scaling limit.
It is expected that close to the tip of the Mott lobe the influence of the $U(1)$ symmetry is strong enough to prevent one from directly (e.g. by using critical exponents) distinguishing between different universality classes from finite system simulations. For other disorder bounds we find $(U/t)_c = 29.44(2)$ for $\Delta = 1.25t$, $(U/t)_c = 29.64(2)$ for $\Delta = 2.5t$, and $(U/t)_c = 30.57(2)$ for $\Delta = 10t$. In the latter case, we directly observed that at the transition point the particle number fluctuations diverge with the system size. Still, even for this relatively strong disorder, finite compressibility can be revealed only by going to very large samples, $L \ge 32$, meaning that the compressibility of the Bose glass phase remains tiny and quickly diminishes away from the critical point.

From our numerical data we extract the phase diagram of the commensurate system shown in Fig.~\ref{fig:phasediagram}. It is clear that the critical disorder strength is always larger than half the energy gap of the homogeneous model by a large margin. Theorem I then immediately leads to the conclusion that the SF-I transition is always to the compressible Bose glass phase, in compliance with Theorem II.

In summary, we have presented an analytic proof of the absence of a direct SF-MI quantum phase transition
in a bosonic system with generic, bounded disorder which resolves a two-decade long controversy. We have shown that any bosonic system with on-site potential disorder is compressible
at the superfluid--insulator critical line and in its  neighborhood.
The quantitative analysis of the Bose-Hubbard model in 3D, based on worm algorithm Monte Carlo simulations, reveals additional details and provides important numerical benchmarks for future experiments. In the limit of small disorder amplitude, the superfluid to insulator transition takes place at the interaction strength $U > U_c^{(0)}$, with $U_c^{(0)}$ being the Mott transition point of the pure system. The overall picture is that disorder first destroys the Mott insulator, converting it into a compressible Bose glass, which is guaranteed to happen (and presumably happens exactly at the condition)
when the bound on the disorder strength reaches the value $\Delta=E_{\rm g/2}(U)$, with $E_{\rm g/2}(U)$ the half-width of the Mott gap in the pure system.
The critical bound on the disorder, $\Delta_c(U)$, corresponding to the onset of disorder-stimulated superfluidity, satisfies relation (\ref{condition}), which guarantees the above-mentioned
picture. Thus, disorder is relevant to the SF-I criticality in all experimentally accessible dimensions.

The {\it theorem of inclusions} is readily generalized to an arbitrary transition in a disordered system between phases A and B, and states that 
one can always find arbitrarily large inclusions of the competing phase on either side of the transition line. The SF-I case with diagonal disorder considered here is one
example; the generic superfluid--Mott(or checkerboard)-glass transition in a particle-hole symmetric system with off-diagonal disorder is another one (cf.~\cite{PS_04, Giamarchi_2001}).
 Whether rare regions are relevant to the nature of the phases A or B remains model specific and is not part of the Theorem. 

We are grateful to Peter Weichman for valuable discussions.
This work was supported by the National Science Foundation
under Grants Nos. PHY-0653183, the DARPA OLE program, the Swiss National Science Foundation and the Aspen Center of Physics. Simulations were performed on the Brutus cluster at ETH Zurich and use was made of the ALPS libraries for the error evaluation~\cite{ALPS}.


\begin{thebibliography}{99}
\bibitem{conjecture} It was conjectured (see Refs.~\cite{Weichman1,Weichman2} for more details) that the  MI-BG transition is exactly at the upper boundary of the theorem, $\Delta_{\rm MI}(U) = E_{\rm g/2}(U)$.
\end{thebibliography}
\end{document}